# Diffusive and Superdiffusive Motion of Sorbates in Carbon Nanotubes

Shreyas Y. Bhide[1], Debashree Ghosh[1], S. Yashonath*[1] and G. Ananthakrishna[2]

[1]Solid State and Structural Chemistry Unit, [2]Materials Research Centre, Indian Institute of Science, Bangalore-560012, India, *email : yashonath@sscu.iisc.ernet.in* (First submitted on Dec, 5, 2003, revised 21st June, 2004)

Recent studies have demonstrated that physical properties of materials at nano length scales are significantly different from that of the bulk.[1] Carbon nanotubes which have attracted much attention in recent times fall into this category.[2] Here we investigate the transport property of spherical sorbates confined to single walled carbon nanotubes (SWNTs) as a function of the size of the sorbates, $\sigma_{gg}$, using molecular dynamics (MD) simulations. The motion is diffusive for small $\sigma_{gg}$, crossing over to superdiffusive motion at intermediate sizes and eventually to ballistic motion as it approaches the diameter of the nanotubes.

Single walled carbon nanotubes are perfect graphene sheets rolled over to form a cylindrical tube having an uniform diameter and smooth walls. Their nano sized dimension coupled with closed topology and helicity are known to be responsible for some remarkable properties that have potential applications. For instance, the elastic modulus of the nanotubes is in the range of 1000 GPa which should find important industrial applications.[3] Recent calculations of self and transport diffusivities in nanotubes show that they are several orders of magnitude larger than in zeolites.[4] Thus, understanding transport properties of sorbates within nanotube is important for applications such as storage and separation of mixtures.

Previous studies of transport within confined media (e.g., other porous solids such as zeolites) have shown some interesting findings. For example, the self diffusivity, $D$, of a sorbate within a porous solid is seen to exhibit a maximum as a function of the size of the sorbate, the levitation effect (LE). The dependence of $D$ on the size of the diffusant, $\sigma_{gg}$, may be divided into two distinct regimes[5] : the *anomalous regime* occurs when the size of the diffusant, $\sigma_{gg}$, is comparable to that of the diameter of the pore, $\sigma_w$ (defined as the distance between the centers of diagonally opposite carbon atoms) and the commonly observed *linear regime* where $\sigma_{gg}$ is significantly smaller than the void radius $\sigma_w$. In the linear regime $D \propto 1/\sigma_{gg}^2$. The LE is an universal behaviour seen in all porous solids irrespective of the nature of the pore geometry and topology.[6] A dimensionless levitation parameter $\gamma$, characterizing the two regimes, is defined as a ratio of the distance at which the optimum interaction to the void radius[5] occurs. $\gamma \sim 1$ in the anomalous regime and much less than unity for the linear regime. In molecular dynamics simulations with Lennard-Jones (LJ) interaction, $\gamma = 2^{7/6}\sigma_{gh}/\sigma_w$. Here $\sigma_{gh}$ and $\sigma_w$ are the guest-host LJ interaction parameter and void diameter respectively.

Studies in zeolites till date have displayed *only diffusive behavior for all sizes of the diffusant or single file diffusion at high loadings in some zeolites*. However, there are several important differences between porous hosts like zeolites and carbon nanotubes, for instance, smooth walls and uniform diameter. In carbon nanotubes, when the size of the diffusant $\sigma_{gg}$ is much smaller than $\sigma_w$, the system is three dimensional. However, when $\sigma_{gg} \approx \sigma_w$, there is essentially one degree of freedom. These differences could be expected to affect the motion within the nanotube. In fact, recent investigations on methane and ethane[7] by us as well as the simulations of $H_2$ and $CH_4$ by Skoulidas et al[4] and Ne and Ar by Ackerman et al[10] corresponding to different effective $\sigma_{gg}$ and $\sigma_w$ reveal that there exist a significant difference in the mobility of these molecules in zeolites compared to carbon nanotubes. These results suggest that the size effect is an important factor in determining the transport properties of sorbates through nanotube as well.

Here we report the results of our study on the influence of size of the sorbates on the transport properties through carbon nanotubes using MD simulations. Detailed investigations carried out here show that diffusive motion gradually changes over to ballistic motion with increase in $\sigma_{gg}$.

*Rigid and flexible carbon nanotube* MD simulations were carried out in the microcanonical ensemble[8] at 200K for various sorbate sizes $\sigma_{gg}$ = 1.8, 2.0, 3.0, 7.0Å in (9,9) carbon nanotubes of length 49.29 Å and $\sigma_w$ = 12.25 Å. $N$ (256) and $N_h$ (16 x 16 x 720 = 184320) are the number of sorbates and host atoms at a loading of one sorbate per channel. 2ns of equilibration and a production run of 20ns for all sizes except for $\sigma_{gg}$ = 7.0 Å where the run length is 100ns. The potential of Hummer et al[9] has been used for flexible nanotube simulations. For more details see Supporting Information.

Table 1 Exponent values of calculated from MD.

| $\sigma_{gg}$(Å) | $\sigma_{gh}$(Å) | $\gamma$ | $\alpha_{MD}$ |
|---|---|---|---|
| 1.8 | 1.8 | 0.32 | 0.95 |
| 2.0 | 2.7 | 0.49 | 1.30 |
| 3.0 | 3.2 | 0.59 | 1.64 |
| 7.0 | 5.2 | 0.95 | 1.94 |

A log-log plot of MSD of the sorbates (averaged over 256 particles and MD trajectories) is shown in Fig. 1 for different $\sigma_{gg}$ values. The MSDs can be fitted to $u^2(t) \sim t^\alpha$. The exponent values have been evaluated using the MSD data in the range 100-1000ps where the statistics is good. This reveals that the transport is essentially diffusive for the smallest size, $\sigma_{gg}$ = 1.8Å with $\alpha$ = 1.03. We also find the temperature dependence of the diffusion constant to be *non-Arrhenius*. Note that the motion is ballistic at short times for small $\sigma_{gg}$, as it should be. Table 1 shows that $\alpha$ *increases from 1 to 2 with size* indicating that the extent of the ballistic transport increases with size and eventually the sorbate behaves like a free particle[11]. Further, flexible carbon nanotube simulations at the same temperature were also carried out and the *trend* seen in Table 1 remains unchanged with a marginal decrease in the exponent value[11].

Some insight into the changes in the dynamics reflected in the increasing superdiffusive behavior can be obtained by studying the nature of the trajectories for various sizes and the increasing superdiffusive behavior can be obtained by studying the nature of the trajectories (projection onto the xy-plane Fig.2), for various sizes and initial conditions. At small sizes ($\sigma_{gg}$ = 1.8 Å), it is clear that the sorbate motion predominantly covers the whole volume within the nanotube as shown in Figs. 2a. With increase in size to 3.0 Å, they are confined to the vicinity of the walls of the nanotube (Fig. 2b). In both of these cases, the motion is largely random or chaotic. This may be seen when the trajectory is projected onto the xy-plane. With further increase in size, the sorbate motion gradually is restricted to an annular region of increasingly smaller radius close to the channel axis. These are shown in Figures 2c and 2d on an expanded scale for size $\sigma_{gg}$ = 7Å. More importantly, the motion is less random or more regular. Such a behavior is seen when the size of particle is comparable to the size of the free diameter of the nanotube (ie.,γ=0.95, corresponding to the anomalous regime of the LE). For this size, the trajectory is nearly one dimensional. The motion unlike that seen in Fig.2a is highly regular leading to well defined patterns. In general, for $\sigma_{gg}$ = 7.0Å, we found that the geometrical structure of the trajectories are sensitive to the initial conditions illustrated for two sets in Figs. 2c and 2d. Figure 2c displays a fan like structure with dominant streaks similar to those seen in the Poincaré maps of Hamiltonian systems. However, we stress that the MSDs obtained here are averages over various initial conditions provided by different particles. For $\sigma_{gg}$ = 7.0Å, for which α =1.94 for 20ns run, we carried out a 100ns (0.1 μs) long run to verify that these results remain unchanged. The value of the exponent from the 100ns run is 1.975, slightly higher than that obtained from the 20ns run suggesting that the run lengths are adequate. We also carried out calculations for 3 guests/channel for $\sigma_{gg}$ = 7Å and found that the superdiffusive motion persists with a marginally lower value of α=1.90.

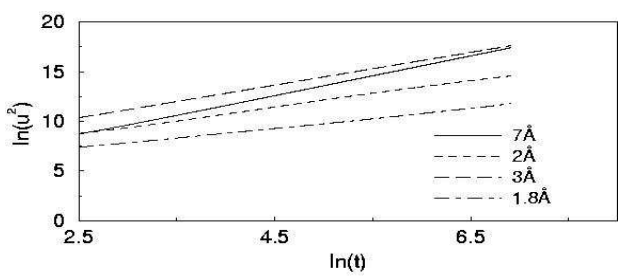

Figure 1 Log-log plot of u²(t) vs *t* for 1ns duration varying over three orders of magnitude for different sizes of the sorbates calculated from MD. $\sigma_{gg}$=1.8 (solid line), 2.0 (dotted), 3.0(dashed) and 7.0 Å(long dashed). The time is in units of ps and MSD in Å².

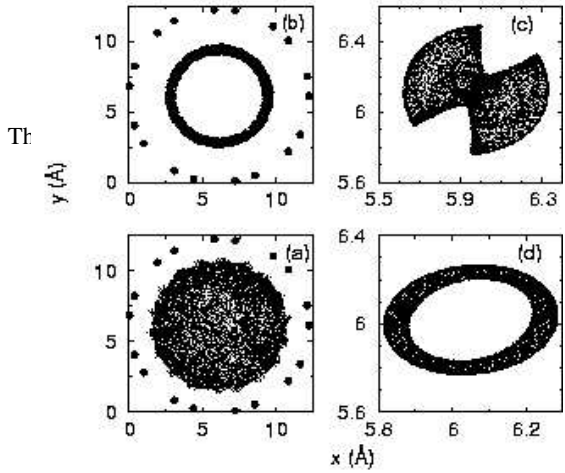

Figure 2 Projections of various types of trajectories for various $\sigma_{gg}$ onto the *xy*-plane. a: $\sigma_{gg}$ = 1.8Å, b: $\sigma_{gg}$ = 3.0Å. Black outer dots represent the atoms of the carbon nanotube. c and d: Fan and ring like structures on an *expanded scale* for $\sigma_{gg}$ = 7.0Å for two different initial conditions.

To understand the crossover from diffusive to ballistic transport, we calculate the mean squared force, $<F_z^2>$, experienced by the sorbate molecule as a function of γ. As can be seen from Fig. 3, the decreasing nature of $<F_z^2>$ with γ signals the change in the nature of the dynamics from a diffusive to superdiffusive nature, ie., $<F_z^2> \to 0$ as $\gamma \to 1$ with increasing size ($\sigma_{gg}$ approaching the free diameter of the nanotube) underscoring the cause of the ballistic motion for $\sigma_{gg}$ =7.0 Å. This should be contrasted with the results obtained in other microporous solids such as zeolites. In zeolites like NaY, NaA or VPI-5, the mean squared force shows a minimum[5] when γ ≈ 1. However, the motion *always remains diffusive*. Actually, zeolites have a nonuniform diameter throughout. Thus, though, $<F_z^2>$ is smaller for particles of size similar to that of the pore or window, force cancellation occurs only at the window connecting the cages. In contrast, as carbon nanotubes *have uniform pore diameter and smooth walls*, the frictional force experienced by the sorbate nearly vanishes everywhere along the tube for large $\sigma_{gg}$ rendering the motion ballistic.

A recent simulation by Skoulidas et al[4] report self and transport diffusivities of $H_2$ and $CH_4$. They find that the self and transport diffusivities of $H_2$ are *higher in the narrower (6,6) nanotube compared to the (10,10) nanotube* by about a factor of 5 and 10 respectively. While these authors explain this on a qualitative reasoning of increasing smoothness[4,10] of narrower tubes, a quantitative explanation of this observation in terms of the levitation parameter γ is easy. An estimate of γ for $H_2$ in (6,6) nanotube is approximately 0.78 compared to relatively small value of ~0.47 in (10,10) nanotube. From Fig. 3, it is evident that $<F_z^2>$ is close to zero for γ = 0.78 and hence lies in the anomalous regime, consequently a higher diffusivity in the (6,6) nanotube. In contrast, for γ = 0.47, $<F_z^2>$ is substantial. (Note that γ is to a large extent independent of the system as it is the ratio of the interaction parameters to the pore size.) However, Ackerman et al[10] report that D for Ne is lowest for (8,8) followed by (10,10) and then by (12,12). More interesting is the non-monotic change in D for Ar, namely, D(8,8)< D(10,10) >D(12,12). Clearly, smoothness alone cannot explain these results. However, they can be understood in the non-monotonic dependence of D on $\sigma_{gg}$ (and therefore through Eq. $\gamma = 2^{7/6}\sigma_{gh}/\sigma_w$ see Figs. 6 and 9 of Ref. 5). Our inference is that D for the case of Ne is in the linear regime where $D_s \propto 1/\sigma_{gg}^2$. For the case of Ar, lies in the crossover from the linear to the anomalous regime. Sokhan et al[12] report viscosity of methane in nanotubes of differing diameters through NEMD simulations.

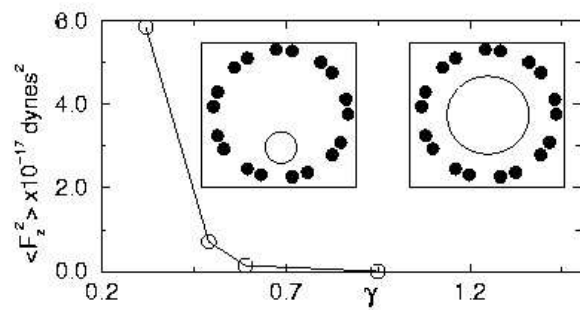

Figure 3 Mean squared force along the nanotube obtained from MD versus the levitation parameter γ. Insets show schematic diagrams for a small and a large sized sorbate in the nanotube.

In summary, our study demonstrates a crossover from a diffusive behavior (with α≈1) at small sizes (γ < 0.5) of the sorbate to superdiffusive nature (1< α < 2) with increase in size eventually approaching ballistic motion when the sorbate size is similar to that of the host. The crossover has been explained quantitatively : $<F_z^2>$ decreases with increase in the levitation parameter (eventually vanishing as γ → 1 (Fig. 3)). The superdiffusive nature is not suppressed even at higher loadings (at least upto 44% of saturation loading). Finally, flexible nanotube simulations for $\sigma_{gg}$ = 7Å at the same temperature yield marginally lower exponent[11]. Indeed, even at a relatively higher temperature of 400K, yielded[11] α = 1.68. More detailed simulations to understand the contributions from different factors to the observed behavior are in progress.

Acknowledgement : Authors wish to thank DST, New Delhi for financial support.

(1) Terrones M. *et. al.*, *Top. Curr. Chem.* **1999.** *199*, 189.
(2) Ajayan, P.M., *Chem. Rev.* **1999**, *99*, 1787.
(3) Yu M.F. *et. al*. *Phys. Rev. Lett.***2000**, *84*, 5552; Tans S. J. *et. al.*, *Nature,* **1997**,*386*, 474.
(4) Skoulidas *et al.*, *Phys. Rev. Lett.,* **2002**,*89*, 185901.
(5) Yashonath, S. and Santikary, P., *J. Phys. Chem.* **1994**, *98*,6368
(6) Bandyopadhyay, S. and Yashonath, S.,*J. Phys. Chem.* **1995**, *99*, 4286.
(7) Bhide, S.Y. and Yashonath, S., *J. Chem. Phys.* **2002**, *116*, 2175; *J. Phys. Chem. A*, **2002**, *106*, 7130.
(8) Allen, M.P. and Tildesley, D. *Computer Simulation of Liquids,* Clarendon Press, Oxford, 1986.
(9) Hummer, G, Rasaiah, J.C., and Noworyta, J.P., *Nature,* **2001**, *414,* 188.
(10) Ackerman D.M. et al., *Mol. Sim.* **2003**, *29*,677.
(11) Bhide, S.Y.; *Diffusion of Polyatomic and Monatomic Sorbates in Zeolites and Microporous Solids : Molecular Dynamics and Lattice Gas Studies*, Ph.D. Thesis, Submitted to Indian Institute of Science, Bangalore, India, December 2002.
(12) Sokhan, V.P., Nicholson, N., Quirke, N.,*J. Chem. Phys.*, **2002**,*117*, 8531.

Abstract

Molecular dynamics simulations of sorbates of different sizes confined to the interior of carbon nanotubes are reported. The mean squared displacement shows gradual change from diffusive for small sorbates to superdiffusive for intermediate sized-sorbates to ballistic for sizes comparable to the channel diameter. We show that this crossover behaviour can be understood on the basis of a gradual decrease of the x-y component of the force with the levitation parameter. The analysis can also help to rationalize some recently published results.

Table of Contents

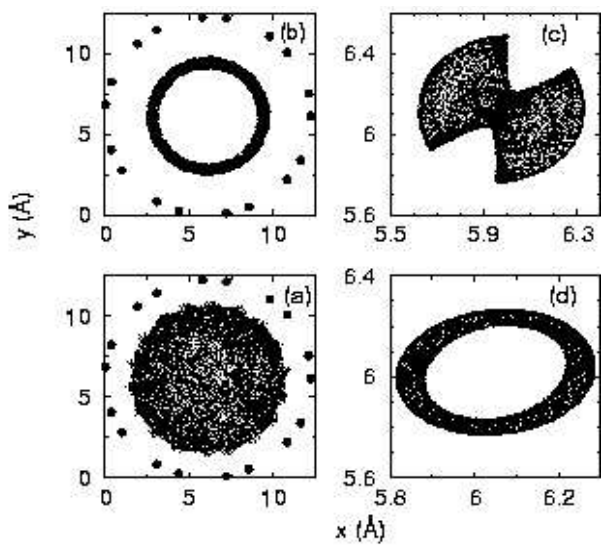